\documentclass[aps,prb,10pt,twocolumn,amsmath,
               amssymb,showpacs,superscriptaddress]{revtex4-1}

\usepackage{graphicx}% Include figure files
\usepackage{bm} % bold math

\def\phm{\phantom{$-$}}

\def\beq{\begin{equation}}
\def\eeq{\end{equation}}
\def\z2{$\mathbb{Z}_2$}
\def\cro{Cr$_2$O$_3$}

\def\e{{\mathcal E}}
\def\eb{{\bm{\mathcal E}}}
\def\bb{{\bf B}}
\def\fb{{\bf F}}
\def\hb{{\bf H}}
\def\pb{{\bf P}}
\def\mb{{\bf M}}

\def\ub{{\bf u}}
\def\A{{\bf A}}
\def\alphab{{\boldsymbol \alpha}}
\def\alphalatb{{\boldsymbol \alpha}^{\rm latt}}

\def\k{{\bf k}}

\def\ket#1{\vert#1\rangle}
\def\bra#1{\langle#1\vert}

\def\wt#1{\widetilde{#1}}

\def\im{\mathrm{Im}\,}

\newcommand{\equ}[1]{Eq.~(\ref{eq:#1})}
\newcommand{\eqs}[2]{Eqs.~(\ref{eq:#1}) and (\ref{eq:#2})}
\newcommand{\equa}[1]{Equation~(\ref{eq:#1})}
\begin{document}

%===========================%
% TITLE PAGE                %
%===========================%
\title{Full magnetoelectric response of \cro\ from first principles}

\author{Andrei Malashevich}
\email{andreim@civet.berkeley.edu}
\affiliation{
Department of Physics, University of California,
Berkeley, California 94720, USA}
\affiliation{Materials Sciences Division, Lawrence Berkeley National
  Laboratory, Berkeley, CA 94720, USA}

\author{Sinisa Coh}
\affiliation{
Department of Physics, University of California,
Berkeley, California 94720, USA}
\affiliation{Materials Sciences Division, Lawrence Berkeley National
  Laboratory, Berkeley, CA 94720, USA}

\author{Ivo Souza}
\affiliation{
Centro de F\'{\i}sica de Materiales (CSIC) and DIPC,
Universidad del Pa\'{\i}s Vasco,
20018 San Sebasti\'an, Spain and
}
\affiliation{
Ikerbasque Foundation, 48011 Bilbao, Spain}

\author{David Vanderbilt}
\affiliation{
Department of Physics \& Astronomy, Rutgers University,
Piscataway, New Jersey 08854, USA}

\date{\today}
\begin{abstract}
  The linear magnetoelectric response of Cr$_2$O$_3$ at zero
  temperature is calculated from first principles by tracking the
  change in magnetization under a macroscopic electric field. Both the
  spin and the orbital contributions to the induced magnetization are
  computed, and in each case the response is decomposed into lattice
  and electronic parts. We find that the transverse response is
  dominated by the spin-lattice and spin-electronic contributions,
  whose calculated values are consistent with static and optical
  magnetoelectric measurements.
  In the case of the longitudinal response, orbital contributions
  dominate over spin contributions, but the net calculated
  longitudinal response remains much smaller than the experimentally
  measured one at low temperatures. We also discuss the absolute
  sign of the magnetoelectric coupling in the two time-reversed
  magnetic domains of Cr$_2$O$_3$.

\end{abstract}
\pacs{75.85.+t,75.30.Cr,71.15.Rf,71.15.Mb}% PACS
% 75.85.+t Magnetoelectric effects, multiferroics
% 75.30.Cr Saturation moments and magnetic susceptibilities 
% 71.15.Mb DFT, LDA, GGA etc.
% 75.80.+q Magnetomechanical effects, magnetostriction
% 71.15.Rf Relativistic effects
% 03.65.Vf Phases: geometric; dynamic or topological
\maketitle
%===========================%
% MAIN TEXT                 %
%===========================%

%---------------------------------------------------------------------

\section{Introduction}

There has been a recent resurgence of interest in magnetoelectric (ME)
couplings in solids.\cite{fiebig-jpd05}
Of particular importance is the {\it linear ME effect},
which can occur in insulating materials with
broken inversion and time-reversal symmetries. It can be described by
a response tensor
\beq
\label{eq:alpha-EH}
\alpha^{\e H}_{ij}
=\left(\frac{\partial P_i}{\partial H_j}\right)_{\eb}
=\mu_0\left(\frac{\partial M_j}{\partial \e_i}\right)_{\hb},
\eeq
where $\pb$ is the electric polarization induced by the magnetic field
$\hb$, and conversely $\mb$ is the magnetization induced by the
electric field $\eb$. 

The early milestones in the long history of the linear ME effect include
the original prediction by Dzyaloshinskii
that it should occur in \cro,\cite{dzyaloshinskii-jetp59} and its
observation shortly after, both in
$\mb(\eb)$~\cite{astrov-jetp60,astrov-jetp61} and in
~$\pb(\hb)$~\cite{folen-prl61,rado-prl61} measurements. The ME effect
has since been observed in a large variety of materials, but \cro\
remains one of the best-studied ME compounds. The early
literature is surveyed in the monograph by O'Dell,\cite{odell-book70}
and recent reviews are given in Refs.~\onlinecite{fiebig-jpd05,
eerenstein-nat06,fiebig-epjb09,rivera-epjb09}.

Most of the early theoretical work was phenomenological in character,
making it difficult to assess the dominant mechanisms behind the ME
response. These can be divided into electronic (i.e. frozen-ion)
vs.\ lattice responses on the one hand,\cite{bonfim-aip80} and
spin vs.\ orbital magnetic contributions on the other.\cite{hornreich-pr67}
{\it Ab initio} theory is an ideal tool for unraveling the microscopic
mechanisms of the ME effect in real materials, and the first
calculations started to appear in recent years. The initial focus was
on spin-lattice contributions,\cite{iniguez-prl08,wojdel-prl09} in
part because investigations of related phenomena in multiferroic
materials over the last decade had indicated that spin-lattice effects
are often dominant there.\cite{cheong-nm07}
In reality, however, very little is known about the relative magnitudes
of the various contributions to the ME tensor in typical magnetoelectric
materials.

Evidence for a significant electronic ME response in \cro\ came
from optical measurements at frequencies above the lattice resonances:
in a series of milestone
experiments,\cite{pisarev-pt91,krichevtsov-jpcm93,krichevtsov-prl96}
Pisarev, Krichevtsov, and collaborators observed
optical effects governed
by an effective ME tensor $\alphab(\omega)$, and found it to be comparable
to the static ME coupling.
Regarding the distinction between spin and orbital couplings (e.g.,
how much of the $\eb$-field induced magnetization comes from spin
moments versus orbital currents), it is probably rather difficult to
separate them experimentally due to the weakness of the ME effect in
known ME materials.
Investigation of the orbital contribution to the ME response is
however interesting in its own right.  In particular, it was
recently established that $\mathbb{Z}_2$ topological insulators
with broken time-reversal symmetry on the surface should display a
quantized electronic orbital ME
response~\cite{qi-prb08,essin-prl09} with a relatively large quantum
($\alpha=24.3$~ps/m in SI units).
This result further suggests that large orbital ME responses can 
in principle be achieved even in generic (non-topological) insulators
with strong spin-orbit coupling without any constraint
on surface preparation.\cite{coh-prb11}

In this paper, we carry out a thorough first-principles investigation
of the linear ME effect in the paradigmatic system \cro.  We compute
the full static response, including on the same footing all four basic
contributions: spin-lattice, spin-electronic, orbital-lattice, and
orbital-electronic.  This completes the programme initiated in
Refs.~\onlinecite{iniguez-prl08} and \onlinecite{bousquet-prl11},
where some but not all of them were evaluated. As in those works, we
shall focus exclusively on the ME response at zero temperature, which
is determined by mechanisms involving the spin-orbit interaction.  

We find that for the response transverse to the rhombohedral axis 
the spin contributions are much larger than the orbital ones. 
The calculated values of the lattice and electronic responses
are in good agreement with both static and optical ME measurements,
as well as with previous calculations.
In the case of the longitudinal response the calculated orbital
contributions are larger than their spin counterparts in both the
electronic and lattice channels. However, as a result of a near
cancellation between the orbital-electronic and orbital-lattice
contributions, the total calculated longitudinal response is
negligibly small.  Thus, the nonzero longitudinal response that
is measured at low temperatures remains unaccounted for.
Some possible reasons for this disagreement will be discussed.

% FIGURE
\begin{figure}
\centering\includegraphics[height=5cm]{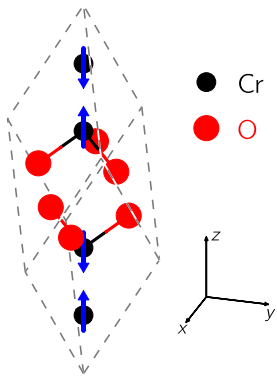}
\caption{(Color online.) Rhombohedral primitive cell of \cro.  The
  arrows indicate the orientations of the {\it magnetic moments} on
  the Cr ions (the {\it spins} on the ions point opposite to the
    arrows). The center of the cell is a center of inversion symmetry
    coupled with time-reversal.
}
\label{fig:cr2o3}
\end{figure}

\section{Preliminaries}

\subsection{Structure of \cro}
\label{sec:prelim_cro}

Chromium (III) oxide (eskolaite)
crystallizes in a corundum-type structure shown in Fig.~\ref{fig:cr2o3},
with two formula units per primitive cell. The magnetic space group
is R$\bar{3}$c$1'$ above $T_N=307$\,K.
Below this temperature \cro\ turns into an antiferromagnetic (AFM)
insulator, with magnetic space group R$\bar{3}'$c$'$. The magnetic
moments on the Cr ions align along the rhombohedral $z$ axis, pointing
up and down in an alternating manner (see Fig.~\ref{fig:cr2o3}). The
magnetic point group is $\overline{3}'m'$, which allows for a diagonal
ME tensor $\alphab$ with two independent components,
$\alpha_{\perp}\equiv\alpha_{xx}=\alpha_{yy}$ (transverse) and
$\alpha_{\parallel}\equiv\alpha_{zz}$
(longitudinal).\cite{newnham-book05}

We note that there are two distinct possibilities for arranging the
magnetic moments in the AFM ground state, related to one another
either by time reversal (i.e., by flipping the magnetic moments on
every Cr ion) or by spatial inversion.  As each of these operations
also flips the sign of $\alphab$, it is important to specify which
configuration is assumed when reporting values for $\alpha_{\perp}$
and $\alpha_{\parallel}$.  Our calculations refer to the configuration
shown in Fig.~\ref{fig:cr2o3}.

\subsection{Formalism and review of previous calculations}

We begin by clarifying issues of units and conventions. \equa{alpha-EH},
which is written in the $(\e,H)$ frame, conforms with the standard
experimental definition of the linear ME tensor, which has units of
ps/m in SI units. Instead, from the point of view of first-principles
theory it is more convenient to work in the $(\e,B)$ frame, where
$\alphab$ has units of vacuum admittance $\sqrt{\epsilon_0/\mu_0}$,
\beq
\label{eq:alpha-EB}
\alpha_{ij}=\left(\frac{\partial{P_i}}{\partial{B_j}}\right)_{\eb}=
\left(\frac{\partial{M_j}}{\partial\e_i}\right)_\bb.
\eeq
The two definitions Eqs. (\ref{eq:alpha-EH}) and (\ref{eq:alpha-EB}) are
related by $\alphab^{\e H}=\mu\alphab$, where $\mu$ is the magnetic
permeability. In the approximation that $\mu/\mu_0\simeq1$, which is
a good approximation for most non-ferromagnetic materials, 
the conversion is trivial, and we shall report the calculated values of
$\alphab$ as though we had computed them in the $(\e,H)$ frame.
For a more detailed discussion, see Sec.~II.A of Ref.~\onlinecite{coh-prb11}.

Let us now discuss how to compute the various contributions to the ME
tensor. To recap, the full response can be decomposed into spin and
orbital parts according to the nature of the induced magnetization in
the $\mb(\eb)$ picture. Each of these can be further decomposed
according to the two basic mechanisms by which the field acts on the
system. 
The electronic part describes the ME response that the
system would have if the ions were held fixed in their equilibrium
positions. The remaining lattice part is associated with the
field-induced ionic displacements.

\subsubsection{Lattice response}
 
We consider first the calculation of lattice couplings. Here
the influence of the applied field ($\bb$ or $\eb$) on the
non-conjugate moment ($\pb$ or $\mb$) is mediated by
internal ionic displacements $\ub$, so that
\beq
\label{eq:alpha_latt}
\alphalatb=\frac{\partial\pb}{\partial\ub}\frac{\partial\ub}{\partial\bb}
=\left(
  \frac{\partial\mb}{\partial\ub}\frac{\partial\ub}{\partial\eb}
\right)^{\rm T},
\eeq
where a summation over the atoms in one crystal cell is implied, and
`T' denotes the matrix transpose.  (In general there may also be a
strain mediated coupling,\cite{wojdel-prl09}
but in \cro\ this contribution vanishes by symmetry, and it will not
be considered further here.)  Optionally, one may also take advantage
of the fact that the displacements induced by the field are mediated
by field-induced forces $\fb$.  For the case of applied electric
field, \equ{alpha_latt} can be rewritten as~\cite{iniguez-prl08}
\beq
\label{eq:alpha_latt_2}
\left(\alphalatb\right)^{\rm T}
=\frac{\partial\mb}{\partial\ub}\frac{\partial\ub}{\partial\fb}
\frac{\partial\fb}{\partial\eb}
=-\Omega\frac{\partial\mb}{\partial\ub}\left(
\frac{\partial^2E}{\partial\ub\partial\ub}\right)^{-1}
\left(
\frac{\partial\pb}{\partial\ub}\right)^{\rm T},
\eeq
where $E$ is the total energy per unit cell and $\Omega$ is the 
unit-cell volume. Here we have made use of the fact that the Born
effective charge tensor can be expressed equivalently as
$(\partial\fb/\partial\eb)^{\rm T}=\Omega\partial \pb/\partial\ub$.
[Alternatively, by invoking the magnetic analog
$(\partial\fb/\partial\bb)^{\rm T}=\Omega\partial \mb/\partial\ub$
we can arrive at this same equation in a different way,
starting from \equ{alpha_latt} for the case of applied magnetic induction.]
Note that the inverse of the
force-constant matrix now appears symmetrically between the magnetic
and electric Born tensors in \equ{alpha_latt_2}.

There are several choices on how to proceed.  One possibility,
following \equ{alpha_latt}, is to relax the structure in the presence
of a small $\bb$ or $\eb$ field, and then compute the
relaxation-induced change in $\pb$ or $\mb$.
Alternatively, \equ{alpha_latt_2} expresses
$\alphalatb$ in terms of three basic quantities (the force-constant
matrix, the Born charges, and their magnetic analogs), all of which
can be computed as changes of various quantities
in response to atomic displacements at vanishing fields.
One can choose to compute such derivatives by finite differences
or by using linear-response techniques available in most
density-functional packages.

\subsubsection{Electronic response}

The calculation of the electronic response $\alphab^{\rm el}$
requires coupling the field $\bb$ or $\eb$ in \equ{alpha-EB} directly
to the electrons, and determining the induced $\pb$ or $\mb$.  In
practice this can be done using either finite-field approaches or
linear-response techniques. 

Of the two contributions, spin-electronic and orbital-electronic, the
latter is the most challenging one to calculate.
A perturbative expression valid for periodic crystals was
recently derived,\cite{malashevich-njp10,essin-prb10} which can
be implemented in the context of density-functional perturbation
theory. In the present work we have opted to calculate the
orbital-electronic response as $\partial\mb^{\rm orb}/\partial\eb$,
using finite electric fields. Another
possibility would be to calculate it as $\partial\pb/\partial\bb^{\rm orb}$,
using a finite orbital magnetic field. The inclusion of
orbital magnetic fields in total-energy calculations of periodic
solids is, however, a challenging problem which has not yet been fully
solved, in spite of some recent
progress.\cite{cai-prl04,gonze-arx11,essin-prb10}

\subsubsection{Review of previous calculations for \cro}

The methods described above were recently used to evaluate the
spin-lattice and spin-electronic parts of ${\boldsymbol\alpha}$.  For
the spin-lattice contribution, \'I\~niguez~\cite{iniguez-prl08}
performed his pioneering calculations following \equ{alpha_latt_2},
while Bousquet {\it et al.}~\cite{bousquet-prl11} used
\equ{alpha_latt}. More precisely, the latter authors performed
structural relaxations in the presence of a fixed Zeeman magnetic
field $\bb^{\rm spin}$ by adding to the Kohn-Sham energy functional a
Zeeman term describing the coupling to the spins.  Furthermore, by
monitoring the linear change in the electronic polarization
${\bf P}^{\rm el}$ under a small field with fixed ions, Bousquet
{\it et al.}  were also able to determine the spin-electronic
response. Thus, out of the four possible contributions to $\alphab$,
only the two spin contributions (lattice and electronic) have
previously been evaluated from first principles for \cro.

\subsection{Computational approach}

Let us now describe the method that we use for calculating the
lattice and electronic ME responses, including in each
case both the spin and the orbital parts of the response.

For the lattice couplings we employ a method similar
to that of Ref.~\onlinecite{iniguez-prl08} but including also the
orbital contribution to $\partial\mb/\partial\ub$ (we
found this to be a more efficient approach than relaxing the lattice
under a finite electric field).  We first compute the Born
effective charges and force-constant matrix using linear-response
techniques,\cite{baroni-rmp01} and from these we find the first-order
field-induced displacements
$\Delta\ub=(\partial\ub/\partial\eb)\cdot\Delta\eb$, where a nominal
field $\Delta\eb$ of $\sim10^9$~V/m is applied along the rhombohedral
axis or in the perpendicular direction.  Displacing the atoms by
$\Delta\ub$, we then determine the induced magnetization $\Delta
\mb^{\mathrm{spin}}+\Delta \mb^{\mathrm{orb}}$. 
The linearity of the magnetization response was checked by
both reducing the magnitude and flipping the sign of $\Delta\eb$.

In order to reduce the computational cost, the spin-orbit interaction is not
included in the linear-response calculations.
This procedure captures the dominant
contributions to $\alphalatb$, i.e., those that are linear in the
spin-orbit coupling strength; we have checked that it produces results
which are almost identical to a calculation in which the spin-orbit
coupling is included at every step.

For a given set of ionic displacements $\Delta\ub$, the orbital
magnetization at $\eb=0$ is calculated under periodic boundary conditions as
\beq
\label{eq:M-zerofield}
\mb^{\mathrm{orb}}=\wt{\mb}^{\mathrm{LC}}+\wt{\mb}^{\mathrm{IC}},
\eeq
where~\cite{xiao-prl05,thonhauser-prl05,ceresoli-prb06}
\begin{eqnarray}
  \label{eq:M_LC}
  \wt{\mb}^{\mathrm{LC}}&=&\frac{e}{2\hbar}\int
  \frac{d^3k}{(2\pi)^3}
  \im\bra{\wt{\nabla}_{\k}u_{n\k}}\times
  H_\k\ket{\wt{\nabla}_{\k}u_{n\k}},\\
  \label{eq:M_IC}
  \wt{\mb}^{\mathrm{IC}}&=&\frac{e}{2\hbar}\int
  \frac{d^3k}{(2\pi)^3}
  \mathrm{Im}\Big[\bra{u_{n\k}}H_\k\ket{u_{m\k}} \nonumber\\
    &&\hspace{2.0cm}
  \bra{\wt{\nabla}_{\k}u_{m\k}}\times\ket{\wt{\nabla}_{\k}u_{n\k}}\Big] .
\end{eqnarray}
Here `LC' and `IC' stand for {\it local circulation} and {\it
  itinerant circulation}, respectively,
$\ket{u_{n\k}}$ is the cell-periodic part of the
Bloch state $\ket{\psi_{n\k}}$, and $H_\k=e^{-i\k\cdot{\bf r}}{\cal
 H}e^{i\k\cdot{\bf r}}$, where ${\cal H}$ is the Kohn-Sham
Hamiltonian of the crystal.
Summations over occupied bands are implied for repeated band indices,
and $\wt{\nabla}_{\k}\equiv(1-\ket{u_{n\k}}\bra{u_{n\k}})\nabla_{\k}$.
In practice the Brillouin-zone integral is replaced by a summation over a 
uniform grid, and $\wt{\nabla}_{\k}$ is evaluated on that grid by
finite differences.\cite{ceresoli-prb06} $\wt{\mb}^{\mathrm{LC}}$ and
$\wt{\mb}^{\mathrm{IC}}$ are separately gauge-invariant, i.e., they
remain unchanged under $k$-dependent unitary transformations among
the occupied states.

Let us now turn to the electronic response, which we calculate as
$\partial\mb/\partial\eb$, taking advantage of the well-established
{\it ab initio} treatment of homogeneous electric fields in periodic
insulators.\cite{souza-prl02} The magnetization
$\mb^{\mathrm{spin}}+\mb^{\mathrm{orb}}$ is determined with and
without an electric field of intensity $\sim10^9$~V/m (using in both
cases the same crystal structure optimized at zero field) in order
to extract the spin-electronic and orbital-electronic ME
couplings.  The evaluation of spin magnetization is straightforward
and here we just mention that, as an additional check, we have
recomputed the spin-electronic coupling using the converse
Zeeman-field approach, finding good agreement between the two methods.

To compute $\mb^{\rm orb}$ at finite $\eb$, we make use of the
following generalization of
Eqs.~(\ref{eq:M-zerofield}--\ref{eq:M_IC}).\cite{malashevich-njp10}
One part is given by the same expression valid at zero field,
\eqs{M_LC}{M_IC}, upon reinterpreting the states $\ket{u_{n\k}}$
therein as field-polarized Bloch states~\cite{souza-prl02} (and ${\cal
  H}$ as the crystal Hamiltonian calculated from the field-polarized
periodic charge density).  To this, an additional contribution of the
form
\beq
\label{eq:M_CS}
\mb^{\mathrm{CS}}=-\frac{e^2}{2\hbar}\eb\int\frac{d^3k}{(2\pi)^3}
\mathrm{Tr}\,\left[\A\cdot\nabla_{\k}\times\A-
\frac{2i}{3}\A\cdot\A\times\A\right]
\eeq
must be added in order to obtain the full orbital magnetization.
Here $\mathbf{A}^{nm}_\k\equiv i\bra{u_{n\k}}\nabla_{\k}\ket{u_{n\k}}$
is the Berry connection matrix; the integrand is a scalar known as
the Chern-Simons 3-form~\cite{qi-prb08,essin-prl09}
(band indices are suppressed). Thus, at $\eb\not=0$ we have,
instead of \equ{M-zerofield},
\beq
\label{eq:M-finitefield}
\mb^{\mathrm{orb}}=\wt{\mb}^{\mathrm{LC}}+\wt{\mb}^{\mathrm{IC}}+
\mb^{\mathrm{CS}}, 
\eeq
and all three terms contribute to the
orbital-electronic ME response. 
The term $\alphab^{\rm CS}$ is purely isotropic and can be
calculated from the valence Bloch states at zero field. Its numerical
evaluation requires a smooth gauge in $k$~space, and this can be
achieved by mapping the valence bands onto localized Wannier
functions.\cite{coh-prb11}
 
%---------------------------------------------------------------------

\subsection{Technical details}

The total-energy and linear-response calculations were performed using
the {\sc Quantum-ES\-PRESSO}~\cite{QE-2009} {\it ab initio} code
package, working in a fully relativistic framework where the
spin-orbit interaction is included in the atomic pseudopotentials.
We employed Troullier-Martins norm-conserving pseudopotentials,
\cite{troullier-prb91} which in the case of Cr included the
semi-core $3s$ and $3p$ states in the valence.

\begin{table}
  \caption{\label{tab:struct}
    Calculated and experimental structural parameters of
    Cr$_2$O$_3$ in the antiferromagnetic phase: rhombohedral lattice parameter $a$, rhombohedral
    angle $\alpha$, and 
    Wyckoff positions of the Cr ions ($4c$ orbit) and O ions ($6e$ orbit).}
\begin{ruledtabular}
\begin{tabular}{lcccc}
& $a$ (\AA) & $\alpha$ (deg) &\multicolumn{2}{c}{Wyckoff positions}\\
&           &              & Cr & O\\
\hline
PBE (This work)                     & 5.415 & 54.45 & 0.1541 & 0.0597\\
LDA (Ref.~\onlinecite{coh-prb11})   & 5.322 & 53.01 & 0.1575 & 0.0690\\
Expt. (Ref.~\onlinecite{hill-mmm10})& 5.358 & 55.0  & 0.1528 & 0.0566\\ 
\end{tabular}
\end{ruledtabular}
\end{table}

The wavefunctions in the
solid were expanded in plane waves with an energy cutoff of 250\,Ry
for structural relaxations and linear-response calculations and
150\,Ry for orbital magnetization calculations.  The Brillouin zone
was sampled on a $4\times4\times4$ Monkhorst-Pack mesh for most
self-consistent-field (SCF) calculations. While this mesh
density produced converged values for the spin-lattice and
spin-electronic ME contributions, the two orbital contributions
converged more slowly with $k$-point sampling [this is probably
related to the finite-differences representation of the covariant
derivatives in \eqs{M_LC}{M_IC}].  After testing several grid
densities, we concluded that a $7\times7\times7$ mesh gave
sufficiently converged values.

As noted in Ref.~\onlinecite{iniguez-prl08}, the computation of
ME couplings demands a very tight tolerance on the convergence of the
self-consistent field loop.
We therefore used rather stringent convergence thresholds, of the order
of $10^{-11}$--$10^{-12}$\,Ry in the total energy.
In order to reach this level of
convergence in a reasonable number of steps with {\sc Quantum-ES\-PRESSO},
we found it useful to use the Thomas-Fermi charge mixing
scheme,\cite{raczkowski-prb01} by setting the input variable
{\tt `mixing\_mode'} to {\tt `local-TF'}. We also found that the speed
of convergence of the calculations with a finite electric field was
improved by increasing the field gradually from zero in small steps.

The exchange-correlation potential was described within the
generalized-gradient approximation (GGA) using the
Perdew-Burke-Ernzerhof (PBE) parametrization.\cite{perdew-prl96} This
choice was made after having optimized the structure using both the
local-density approximation (LDA) and PBE, and finding that the latter
produced structural parameters in better agreement with experiment
(see Table~\ref{tab:struct}).
In particular, LDA underestimates the unit-cell volume by 7.3\% while
PBE overestimates it by only 1.7\%.  We note that the authors of
Refs.~\onlinecite{iniguez-prl08} and \onlinecite{bousquet-prl11} used
LDA+U with the experimental cell volume enforced.
As for the magnetic structure, the staggered spin moments
on the Cr atoms have a value of $2.7$\,$\mu_\mathrm{B}$/atom,
for a sphere integration radius of $1.3$~\AA.
This is in good agreement with the LDA+U value reported
in Ref.~\onlinecite{iniguez-prl08}.

%%%%%%%%%%%%%%%%%%%%%%%%%%%%%%%%%%%%%%%%%%%%%%%%%%%%%%%%%%%%%%%%%%%%%%%%%%%%%

\section{Results}

%--------------------------------------------------------------------------
\subsection{Contributions to the ME response}
\label{sec:res_contribs-dv}
%--------------------------------------------------------------------------

\begin{table}
  \caption{\label{tab:alpha_both}
    Calculated contributions to the magnetoelectric tensor components
    $\alpha_{\perp}$ and $\alpha_{\parallel}$ in \cro\ .
    Columns (rows) show the spin and orbital
    (electronic and lattice) contributions.
    [The results from previous calculations are indicated in parentheses.]}
\begin{ruledtabular}
\begin{tabular}{lccccccc}
& \multicolumn{3}{c}{$\alpha_{\perp}$ (ps/m)} & \phantom{sp} &
  \multicolumn{3}{c}{$\alpha_{\parallel}$ (ps/m)} \\
& Spin & Orb. & Total & & Spin & Orb. & Total \\
\hline
Elec.      & 0.26 & $-$0.014 & 0.25 & & 0.0007 & $-$0.009  & $-$0.008 \\
           & (0.34\footnotemark[1]) & & & & (0\footnotemark[1]) & & \\
Latt.      & 0.77 & \phm0.025 & 0.80 & & 0.0026 & \phm0.008 & \phm0.011 \\
           & (1.11\footnotemark[1]) & & & & (0\footnotemark[1]) & & \\
           & (0.43\footnotemark[2]) & & & & (0.00\footnotemark[2]) & & \\
Total      & 1.03 & 0.011 & 1.04 & & 0.003 & $-$0.001 & \phm0.002 \\
\end{tabular}
\footnotetext[1]{Ref.~\onlinecite{bousquet-prl11}.}
\footnotetext[2]{Ref.~\onlinecite{iniguez-prl08}.}
\end{ruledtabular}
\end{table}

The main results of our calculations are presented in
Table~\ref{tab:alpha_both} together with results from previous
theoretical works, given in parentheses.  Let
us first analyze the transverse ME response. The magnitude of the
calculated static value, $|\alpha_\perp|=1.04$~ps/m, agrees well with
the most reliable measurements, which range from 0.7 to
1.6~ps/m.\cite{kita-jap79,wiegelmann-fe94} The spin-lattice
contribution accounts for about 75\% of that value, with the remaining
25\% coming mostly from the spin-electronic response, while the two
orbital contributions are negligible (less than 2\%).  The values we
obtain for the individual contributions $\alpha_\perp^{\rm latt}$ and
$\alpha_\perp^{\rm el}$ agree well with those calculated in
Ref.~\onlinecite{bousquet-prl11} using the converse Zeeman-field
approach.

In the case of the longitudinal response, the relative strengths
of the four contributions are very different. As in previous
calculations,\cite{iniguez-prl08,bousquet-prl11} we find that the
spin contributions to $\alpha_\parallel$ are very small, summing
to only 0.003\,ps/m in our calculation.  This can be understood
as resulting from the extreme stiffness of the magnitude of the
spin moment in a collinear band antiferromagnet, which is also
reflected in the near-vanishing of the spin magnetic susceptibility
$\chi_\parallel$ at $T=0$.\cite{hornreich-pr67}

Experimentally, however, the low-temperature $\alpha_\parallel$ is
found to be about 0.2--0.3\,ps/m.\cite{kita-jap79,wiegelmann-fe94}
This is smaller than $\alpha_\perp$ by a factor of 3 to 6, but
still about two orders of magnitude larger than our theoretical spin
value, suggesting that orbital effects might be responsible for most
of the $\alpha_\parallel$ response.  Indeed, Hornreich and
Shtrikman~\cite{hornreich-pr67} pointed out that a zero-temperature
longitudinal ME response could arise in \cro\ from an
electric-field-induced shift in the $g$ factor of the Cr ions (see
also Ref.~\onlinecite{bonfim-aip80}).  This is an orbital effect that
should be automatically included in the present calculations. In fact,
we do find that our computed orbital-lattice and orbital-electronic
contributions to $\alpha_\parallel$ are nearly an order of magnitude
larger than the corresponding spin contributions.  However, the
orbital-lattice and orbital-electronic contributions individually are
still an order of magnitude smaller than the measured value.
Moreover, these two contributions have opposite signs, resulting in a
near cancellation of the entire longitudinal response.  Our total
$\alpha_\parallel$ of 0.002\,ps/m thus remains about two orders of
magnitude smaller than the measured value.

There are several possible explanations for this discrepancy.
First, the theoretical values for the orbital longitudinal
response are quite small, and thus might be especially sensitive to
numerical errors.  However, we have checked $k$-point and self-consistent
convergence carefully, and do not believe this is a major concern.
More serious is the potential dependence on choice of exchange-correlation
potential. 
In particular, within the LDA we found that the orbital-electronic and
orbital-lattice contributions are approximately a factor of 3
larger compared to PBE, although similar cancellation of the two
contributions was observed.
Future work is needed to check the sensitivity of these calculations
to the choice of GGA (adopted here) as opposed to LDA, LDA+U or GGA+U,
hybrid functionals, or other 
orbital-dependent functionals.  Since orbital
currents play a crucial role, the use of current-density functionals
should probably also be explored.
On the experimental side, it would probably be advisable to check
the dependence of the measured value on sample quality, in order
to rule out extrinsic effects associated with defects, surfaces,
contacts, etc.

It is also possible, however, that the experimentally observed
response is dominated by some physics not captured by LDA or GGA
approximations to the exact density functional.  For example, the
strong dependence of $\alpha_\parallel$ upon temperature makes it
clear that thermal fluctuations strongly influence the longitudinal
response.  By the same token, it is possible that quantum spin
fluctuations, already present in the antiferromagnetic state at zero
temperature, may play an important role.  For the time being, we leave
this as an open question.

Before closing this section, we recall that the orbital
ME response can be further decomposed into local circulation
(LC), itinerant circulation (IC), and --~in the case of
the orbital-electronic response~-- Chern-Simons (CS)
contributions, as in \equ{M-finitefield}.
Table~\ref{tab:orb} shows the
breakdown of the full orbital response computed in the present work.
In our previous study of \cro, only the isotropic CS term was
calculated (using LDA rather than GGA).\cite{coh-prb11}
In that work we found the CS term to be $\sim0.01$~ps/m,
about an order of magnitude larger than the presently calculated value.
Further work is needed to determine how the various terms in the ME response
of \cro\, depend on the choice of exchange-correlation potential.
It can be seen that the CS contribution to the orbital-electronic
response is about an order of magnitude smaller than the LC and IC
contributions. Individually, the LC and IC orbital-electronic
contributions are somewhat larger for $\alpha_\parallel$ than for
$\alpha_\perp$, but taken together the opposite is true.
As for the orbital-lattice contributions to $\alpha_\perp$ and
$\alpha_\parallel$, they come mainly from the LC terms.

\begin{table}
\caption{\label{tab:orb}
Decomposition of the calculated orbital ME response of \cro\
(presented in Table~\ref{tab:alpha_both})
into ``local circulation'',
``itinerant circulation'',
and ``Chern-Simons''
contributions coming respectively from
Eqs.~(\ref{eq:M_LC}), (\ref{eq:M_IC}), and (\ref{eq:M_CS}).
}
\begin{ruledtabular}
\begin{tabular}{lll}
$\alphab^{\mathrm{orb}}$ (ps/m) & $\alpha_{\perp}^{\mathrm{orb}}$ &
                                 $\alpha_{\parallel}^{\mathrm{orb}}$ \\
\hline
Electronic           &       &           \\
\quad Local circulation & $-$0.0064 & $-$0.0237  \\
\quad Itinerant circulation & $-$0.0084 & \phm0.0135  \\
\quad Chern-Simons   & \phm0.0012 & \phm0.0012 \\
\quad Subtotal       & $-$0.0136 & $-$0.0090 \\
Lattice              &       &           \\
\quad Local circulation & \phm0.0202 & \phm0.0078     \\
\quad Itinerant circulation & \phm0.0051 & \phm0.0000     \\
\quad Subtotal       & \phm0.0253 & \phm0.0078     \\
Total                & \phm0.0117 & $-$0.0012  \\
\end{tabular}
\end{ruledtabular}
\end{table}
%

%---------
\subsection{Sign of the ME response}
%---------

We now discuss the overall sign of the tensor $\alphab$.
As already mentioned, in \cro\, this sign
depends on the orientation of the magnetic moments (see
Fig.~\ref{fig:cr2o3}).  Experimentally, a single AFM domain can be
stabilized by cooling the sample through the N\'eel temperature in the
presence of parallel (or antiparallel) 
electric and magnetic fields (``magnetoelectric
annealing''), and the spin structure can then be analyzed using
spherical neutron polarimetry.\cite{brown-jpcm02,brown-sss05}

According to Ref.~\onlinecite{brown-jpcm02},
the orientation of the magnetic moments shown
in Fig.~\ref{fig:cr2o3} therein corresponds to a domain annealed with electric
and magnetic fields pointing in the opposite direction along the
rhombohedral axis, provided that arrows in that figure indeed indicate
directions of spin moments rather than magnetizations.
Since the magnetoelectric tensor appears in the free energy in the form
$F_{\mathrm{ME}}=-\alpha_{ij}\e_iH_j$, the domain under consideration
should have negative $\alpha_{\parallel}$ near the N\'eel temperature.
Experimental measurements of magnetoelectric coupling as a function of
temperature\cite{kita-jap79,wiegelmann-fe94} show that
$\alpha_{\parallel}$ changes sign around $100$~K, while
$\alpha_{\perp}$ is negative all the way to $4.2$~K.
Assuming that magnetic domain is determined at high temperatures,
close to the N\'eel temperature, and that magnetic domains remain frozen
upon cooling to $4.2$~K, we can conclude that at $4.2$~K the domain shown
in Fig.~\ref{fig:cr2o3} must have $\alpha_{\perp}>0$ and
$\alpha_{\parallel}>0$.

Our computed signs appear to agree with the experimental
work of Ref.~\onlinecite{brown-jpcm02},
although it was not made entirely clear whether
the signs reported there refer to spins or magnetizations.
Now that first-principles theory is seriously beginning to
confront experiment in the field of magnetoelectric couplings,
we urge closer attention to sign issues in future investigations,
both theoretical and experimental.

\subsection{Comparison to optical measurements}

We now turn to the comparison with existing measurements of the
optical ME tensor $\alphab(\omega)$. As our theory only deals with
static fields, the calculated $\alphab^{\rm el}$ should be thought of
as the $\omega\rightarrow 0$ limit of the purely electronic
optical response (quasistatic limit). This is expected to
approximate reasonably well the measured response at frequencies
between the lattice and electronic resonances and sufficiently far from both.

The ME coupling influences both the transmission and
reflection of light from a magnetoelectric medium, giving rise to
characteristic optical effects which are odd under
time reversal.\cite{bonfim-aip80,hornreich-pr68} While the
propagation of electromagnetic waves inside a ME medium is only
affected by the traceless part of $\alphab$, all tensor components
can in principle be extracted from reflectance measurements,
although in that case the net effect may also have surface-specific
contributions.\cite{krichevtsov-jpcm93} The reflection experiments
of Ref.~\onlinecite{krichevtsov-jpcm93} were carried out using
visible light with a wavelength of 633~nm (1.96~eV), which falls
within the exciton absorption range of \cro, thus precluding
a meaningful comparison with our quasistatic calculations.

We therefore focus on the earlier transmission
measurements,\cite{pisarev-pt91} which used near infrared light of
1156~nm (1.07~eV). The effect that was observed consists of a tilt
away from the crystallographic $\hat{\bf y}$ and $\hat{\bf z}$
directions of the linear polarization of light traveling along
$\hat{\bf x}$. The tilt angle $\phi$ is related to the components of
the optical ME tensor (expressed in Gaussian units) and index of
refraction by\cite{hornreich-pr68,krichevtsov-jpcm93}
\beq
\phi\simeq -\frac{1}{2}
\frac{\alpha_{zz}-\alpha_{xx}}{n_z-n_x}.
\eeq
While an effect which changed sign between time-reversed samples was
clearly observed, a time-even background signal of comparable
magnitude could not be eliminated. The most reliable value,
$\phi=4'\simeq 1.2\times 10^{-3}$~rad, was measured at 220-240~K. As
the absolute value of the linear birefringence was not reported, we
use the value $n_z-n_x =5.8\times 10^{-2}$ quoted in
Ref.~\onlinecite{krichevtsov-jpcm93} for 633~nm, to arrive at
$\alpha_{xx}-\alpha_{zz}\sim 0.12$~ps/m.  The agreement with our
calculated value of 0.26~ps/m is quite satisfactory, given the
experimental uncertainties as well as the limitations in our theory
(namely, the DFT underestimation of the optical gap and the assumed
quasistatic and low temperature limits in the calculation).

We emphasize that, as in the case of the static measurements discussed
earlier, the dominant type of AFM domain present in the samples was
not specified in Ref.~\onlinecite{pisarev-pt91}.  Hence the {\it
  absolute} sign of the measured optical ME coefficient was not
determined.
  It would be interesting to carry out optical and
  static ME measurements on the same single-domain sample at low
  temperatures. This would allow one to extract the {\it relative} sign
  between $\alpha_\perp^{\rm el}$ and $\alpha_\perp^{\rm
    latt}+\alpha_\perp^{\rm el}$, which we predict to be positive.

%%%%%%%%%%%%%%%%%%%%%%%%%%%%%%%%%%%%%%%%%%%%%%%%%%%%%%%%%%%%%%%%%%%%%%%%%%%%%
%
\section{Conclusions}
In summary, we have performed a thorough investigation of the
zero-temperature ME response in \cro\ using first-principles
calculations. We analyzed the lattice and electronic parts of the
response including both spin and orbital magnetization
contributions, being careful to treat all four contributions on an
equal footing. In particular, we treated the orbital response
using the modern Berry-phase theory,
without introducing muffin-tin approximations,
in which orbital currents are computed 
inside spheres around atoms.

We have then compared the calculated values with static and optical
measurements.  Previous calculations, which focused on the spin
contributions, had found an essentially null value for
$\alpha_\parallel$, in disagreement with experiment. We therefore set
out to check whether orbital effects could account for the observed
low-temperature longitudinal response, as had been proposed early on
in the literature.  Our results suggest that this is not the case, as
the calculated orbital responses are very small, consistent with
a scenario of strongly quenched orbital moments. We hope that the
present findings will stimulate further investigations, both on the
experimental and theoretical sides.

Recently we became aware of
concurrent first-principles studies of the orbital ME response in \cro\
\cite{fechner-12} and LiFePO$_4$\cite{scaramucci-arx12} using the
approximation of integrating orbital currents within atom-centered
spheres.

This work was supported by NSF Grant DMR-10-05838 and by
Grant MAT2012-33720 from the Spanish Ministerio de Econom\'{\i}a y
Competitividad. We would like to thank Michael Fechner, Manish
Jain, and Georgy Samsonidze for useful discussions.

\end{document}